\documentclass[conference]{IEEEtran}
\IEEEoverridecommandlockouts
\usepackage[T1]{fontenc}
\usepackage{cite}
\usepackage{amsmath,amssymb,amsfonts}
\usepackage{algorithmic}
\usepackage{graphicx}
\usepackage{textcomp}
\usepackage{xcolor}
\usepackage{comment}
\usepackage{flushend}
\usepackage{orcidlink}
\def\BibTeX{{\rm B\kern-.05em{\sc i\kern-.025em b}\kern-.08em
    T\kern-.1667em\lower.7ex\hbox{E}\kern-.125emX}}
    
\begin{document}

\title{Large-scale characterization of Single-Hole \\Transistors in 22-nm FDSOI CMOS Technology}

\author{\IEEEauthorblockN{Thomas H. Swift\IEEEauthorrefmark{1}\IEEEauthorrefmark{2}\IEEEauthorrefmark{6},
Alberto Gomez-Saiz\IEEEauthorrefmark{1}\IEEEauthorrefmark{3},
Virginia N. Ciriano-Tejel\IEEEauthorrefmark{1}, 
David F. Wise\IEEEauthorrefmark{1},
\\
Grayson M. Noah\IEEEauthorrefmark{1},
John J. L. Morton \IEEEauthorrefmark{1}\IEEEauthorrefmark{2},
M. Fernando Gonzalez-Zalba\IEEEauthorrefmark{1}\IEEEauthorrefmark{4}\IEEEauthorrefmark{5}\IEEEauthorrefmark{6}
Mark A. I. Johnson\IEEEauthorrefmark{1}}
\IEEEauthorblockA{\IEEEauthorrefmark{1} Quantum Motion, 9 Sterling Way, London, N7 9HJ, United Kingdom}
\IEEEauthorblockA{\IEEEauthorrefmark{2}London Centre for Nanotechnology, UCL, London, WC1H 0AH, United Kingdom}
\IEEEauthorblockA{\IEEEauthorrefmark{3}Department of Electrical and Electronic Engineering, Imperial College London, London SW7 2AZ, United Kingdom}
\IEEEauthorblockA{\IEEEauthorrefmark{4}CIC nanoGUNE Consolider, Tolosa Hiribidea 76, E-20018 Donostia-San Sebastian, Spain}
\IEEEauthorblockA{\IEEEauthorrefmark{5}IKERBASQUE, Basque Foundation for Science, E-48011 Bilbao, Spain}
\IEEEauthorblockA{\IEEEauthorrefmark{6} tom.swift@quantummotion.tech, fernando@quantummotion.tech}

}

\maketitle

\begin{abstract}
State-of-the-art quantum processors have recently grown to reach 100s of physical qubits. As the number of qubits continues to grow, new challenges associated with scaling arise, such as device variability reduction and integration with cryogenic electronics for I/O management. Spin qubits in silicon quantum dots provide a platform where these problems may be mitigated, having demonstrated high control and readout fidelities and compatibility with large-scale manufacturing techniques of the semiconductor industry. Here, we demonstrate the monolithic integration of 384 p-type quantum dots, each embedded in a silicon transistor,  with on-chip digital and analog electronics, all operating at deep cryogenic temperatures. The chip is fabricated using 22-nm fully-depleted silicon-on-insulator (FDSOI) CMOS technology. We extract key quantum dot parameters by fast readout and automated machine learning routines to determine the link between device dimensions and quantum dot yield, variability, and charge noise figures. Overall, our results demonstrate a path to monolithic integration of quantum and classical electronics at scale.   
\end{abstract}


\section{Introduction}

Spin qubits hosted in semiconductor quantum dots have been shown to operate above the threshold for fault-tolerance~\cite{xue_quantum_2022} but as technology scales-up to computationally useful levels of complexity, two important challenges arise: (i) I/O management \cite{franke_rents_2019,reilly_challenges_2019} and (ii) qubit variability mitigation\cite{neyens_probing_2024}.

In current I/O approaches, the number of addressing lines that connect room-temperature electronics with the quantum processor hosted below 1~K, scales linearly with the number of qubits. For large-scale systems, this type of scaling is unsustainable due to increased system complexity, reduced reliability, increased heat load, and cost. To manage I/O, frequency-division multiple access (FDMA) can be utilized. FDMA allows multiple qubits to share measurement electronics, however frequency crowding currently limits this approach to 8 qubits per line \cite{george_multiplexing_2017}. Another approach to I/O management is crossbar architectures \cite{bavdaz_quantum_2022}, for which $\mathcal{O}\left (\sqrt{N}\right )$ addressing lines are used to control $N$ qubits. However, this approach places strict limits on qubit variability, which cannot be met with current manufacturing technology, leading to the second challenge
: to minimize manufacturing process variability and consequently improve qubit performance at scale\cite{gonzalez-zalba_scaling_2020}. 

To address this challenge, variability needs to be quantified and traced back to its source within the manufacturing process. For quantum dot (QD) devices, this step involves the development of high-volume and rapid cryogenic testing techniques. Current state-of-the-art approaches utilize cryogenic wafer probers, but these machines are limited to operation temperatures above 1.6~K \cite{neyens_probing_2024}, a temperature higher than the optimal for qubit operation. Alternatively, on-chip multiplexing can be used to access a large number of devices with a small number of lines while remaining operational well below 1~K~\cite{ PaqueletWuetz2020, tosato2025qarpetcrossbarchipbenchmarking}. Previous work demonstrated the design and rapid characterization of a multiplexed array (referred to as a farm) of n-type QDs formed in transistors in a 22-m FDSOI CMOS technology~\cite{thomas_rapid_2023}. Here, we expand previous work to the study of p-type QD devices. We determine device yield, QD electrostatic-parameter variability and charge noise levels across a farm of 384 devices of varying gate lengths and channel widths using fast readout and automated data analysis tools.


\section{DC Transistor Characterization}

\begin{figure}[htbp]
\centerline{\includegraphics[width=0.8\columnwidth]{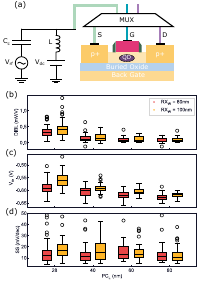}}
\caption{(a) Depiction of transistor structure used for both classical and QD characterization. Box and whisker plots are then shown for (b) DIBL (c) $V_{\mathrm{th}}$ and (d) SS vs transistor gate length (PC$_{\mathrm{L}}$). $V_{\mathrm{th}}$ and SS are measured at $V_{\mathrm{sd}}$ = 100~mV and DIBL is calculated from $V_{\mathrm{sd}}$ = 25, 50 \& 100 mV.}
\label{fig:figure_1}
\end{figure}

The devices measured are p-type transistors manufactured by GlobalFoundries using their 22FDX\circledR~process and have a structure as shown in Fig. \ref{fig:figure_1}~(a). These devices use a conventional epi-silicon body rather than the strained silicon  body (eSiGe) used for typical p-type transistors in 22FDX\circledR. The devices are embedded within a farm,  which allows digitally controlled multiplexed access to one device at a time. The farm contains multiple copies of devices with front gate lengths (PC$_{\mathrm{L}}$) 28-80~nm and channel widths (RX$_{\mathrm{W}}$) 80 \& 100~nm. These short gate lengths are necessary for the formation of QDs. Previous work has shown a strong correlation between the threshold voltage ($V_{\mathrm{th}}$) at room temperature and the first electron voltage ($V_{\mathrm{1e}}$) of a QD at cryogenic temperatures \cite{thomas_rapid_2023}. Here we extend such classical characterization to deep cryogenic temperatures and additionally extract drain-induced barrier lowering (DIBL) and subthreshold swing (SS). The $V_{\mathrm{th}}$ is extracted using a constant current density method in the transistor linear regime, with the DIBL then calculated using the change in $V_{\mathrm{th}}$ as a function of source voltage. All measurements are taken at base temperature ($\sim$~20~mK) of a Bluefors XLD, with on-chip power dissipation raising the sample temperature to around 600~mK as determined using on-chip diode thermometry \cite{noah_cmos_2024}.

\begin{figure}[htbp]
\centerline{\includegraphics[width=0.7\columnwidth]{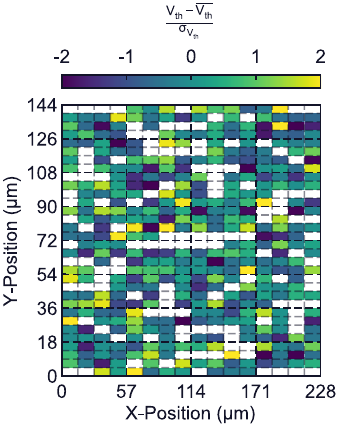}}
\caption{Deviation of threshold voltage from the mean value of each device type across the farm normalized by the standard deviation (white squares indicate other device types that were not measured in this experiment).}
\label{fig:figure_2}
\end{figure}

Fig. \ref{fig:figure_1}~(b)-(d) shows box and whisker plots of the extracted $V_{\mathrm{th}}$, DIBL and SS values for the various measured device types. We found that $|V_{\mathrm{th}}|$ decreases for shorter gate lengths whereas the DIBL increases. This behavior can be explained as the result of short channel effects. As the gate length is reduced, the distance between the source and drain also decreases and becomes comparable to the width of the depletion layer. The channel barrier potential is then influenced by the drain potential, leading to DIBL, which has the effect of reducing $V_{\mathrm{th}}$. The standard deviation of $V_{\mathrm{th}}$ is also observed to increase with decreasing gate length, which could be related to subthreshold peaks of QDs or dopants \cite{wacquez_single_2010,sellier_subthreshold_2007} or could be a consequence of the increased sensitivity to the manufacturing variability of devices with smaller dimensions. The subthreshold swing does not show a dependence on device parameters, with an average value of 16.1~mV/dec which agrees well with literature values of 9.93~mV/dec for 40~nm MOSFETS \cite{yang_quantum_2020} and is above the theoretical minimum of $SS\sim\frac{kT}{q}\ln 10\approx$ 0.1~mV/dec for a chip temperature of 600~mK \cite{sze_physics_2007}. To ensure that there are no systematic deviations in transistor behavior across the farm, the deviation of $V_{\mathrm{th}}$ from the average value for each specific device type is plotted as a function of position within the farm, normalized by the standard deviation. This plot as shown in Fig. \ref{fig:figure_2} demonstrates no clear correlation between the spatial location within the farm and the deviation from the average $V_{\mathrm{th}}$ (white squares in the plot indicate other device types that were not measured in this experiment). 

\section{Quantum Dot Characterization}

Having understood the classical behavior of these transistor and the effect of device dimensions, the next step is to characterize their ability to form QDs. To gain an understanding of the QD structure that is formed, the gate, $V_\text{G}$, and the source-drain voltage $V_\text{SD}$ are varied while monitoring the in-phase response of an LC resonator ($I$) that is sensitive to changes in QD impedance. The voltage at which the first hole is observed to be loaded onto the QD ($V_{\mathrm{1h}}$) is extracted as well as the gate lever arm ($\alpha$) which is a measure of the strength of the electrostatic coupling between the gate electrode and the QD. The lever arm is specifically defined as the ratio of the gate and total capacitance of the QD.

\begin{figure}[htbp]
\centerline{\includegraphics[width=\columnwidth]{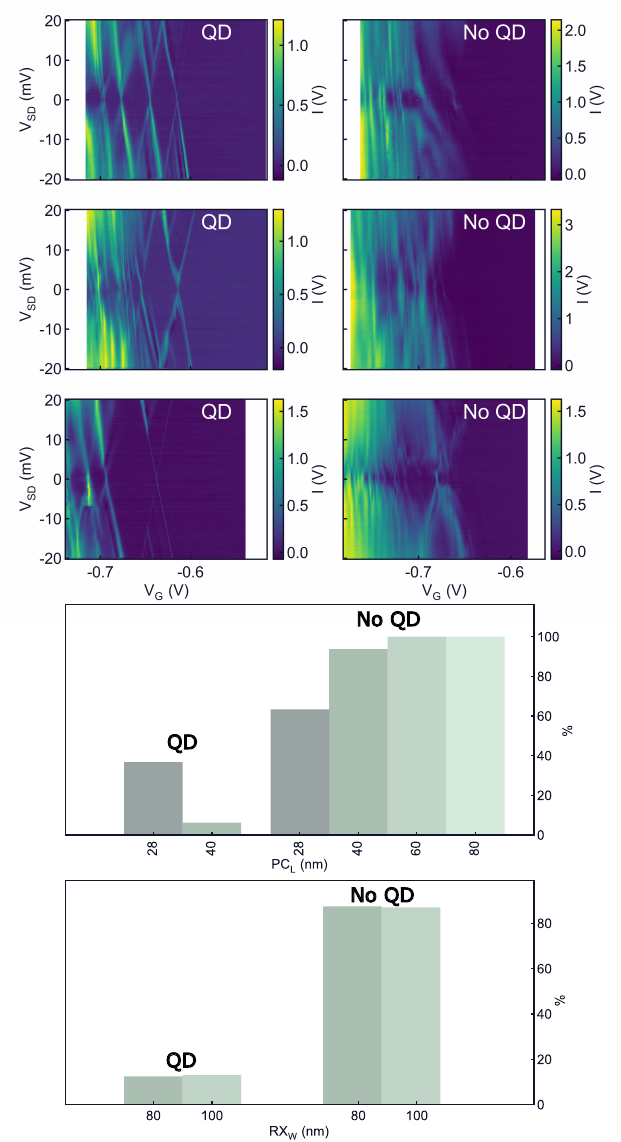}}
\caption{(Top panels) Comparison of Coulomb diamond maps characterized by "Diamondsky" as being from `good' vs `bad' quantum dots (Bottom panels) Percentage of devices being characterized as good or bad depending on their channel width (RX$_\text{W}$) and gate length (PC$_\text{L}$).}
\label{fig:figure_3}
\end{figure}

To investigate the optimal device dimensions, the `QD' vs `No QD' characterization of the "Diamondsky" analysis package \cite{thomas_rapid_2023} is used. This classification was determined by training a convolutional neural network using labeling by domain experts. The top panels of Fig. \ref{fig:figure_3} show examples of devices characterized as `QD' and `No QD'. It is clear from these plots that the observation of a QD in this case is given by clear and sharp edges of the Coulomb oscillations and at least one discernible Coulomb diamond. The bottom panels show the percentage of `QD' or `No QD' depending on the device dimensions. This shows that the most important parameter is PC$_{\mathrm{L}}$ and for PC$_{\mathrm{L}}$ = 28~nm we see $\sim$~30$\%$ good dots compared to the  $<$10$\%$ for the next shortest value of PC$_{\mathrm{L}}$ = 40~nm. 
Overall, the data suggests that the focus of future fabrication runs should be on PC$_\text{L}$ $\leq$ 28~nm. However, it should also be considered that pushing towards the minimum achievable gate length of the process will likely lead to increased device variability, as shown by the increase in standard deviation of $V_{\mathrm{\mathrm{th}}}$ with decreasing PC$_\text{L}$ as shown in Fig. \ref{fig:figure_1}.

\section{Charge Noise Investigation}

\begin{figure}[htbp]
\centerline{\includegraphics[width=\columnwidth]{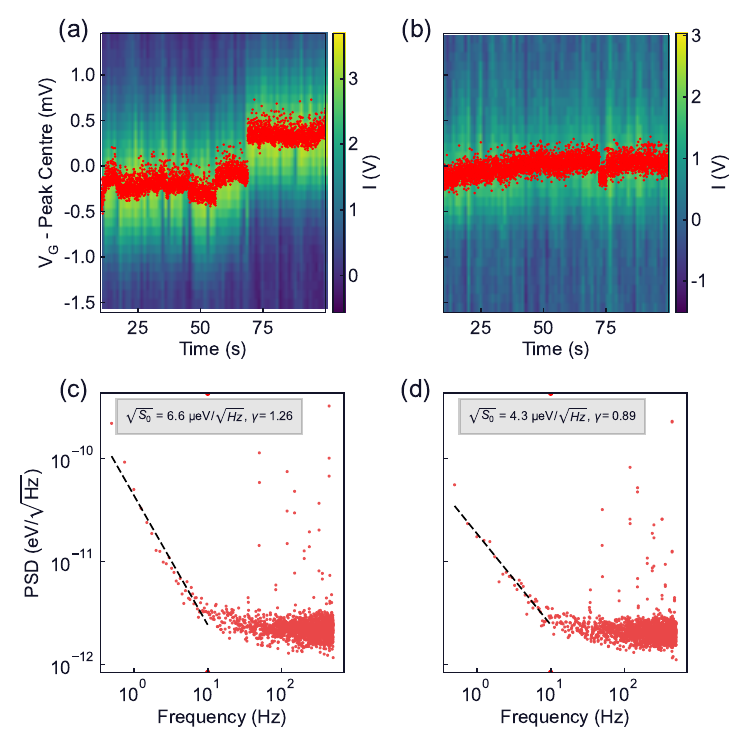}}
\caption{(a)-(b) In-phase signal ($I$) against time and gate voltage for two different device over 100 seconds (c)-(d) Power spectral density calculated from peak position data.}
\label{fig:figure_4}
\end{figure}

As well as characterizing the quality of the QDs formed by single-hole transistors in the process, it is also important to understand the environment within which the QD is situated, in particular the charge noise. Both charge and spin qubits are negatively affected by charge noise so it is important to characterize and understand the nature of charge noise and two-level fluctuators in the process to inform future process optimization. Fig. \ref{fig:figure_4} (a)~\&~(b) show the measurement data from a peak tracking measurement, specifically the in-phase signal ($I$) as a function of gate voltage around the average Coulomb peak center over a period of 100 seconds. Red dots identify the fitted peak position, overlaid on top of the physical signal. Fig. \ref{fig:figure_4} (c)~\&~(d) then shows the associated power-spectral-density of (a)~\&~(b) calculated using Welch's method \cite{welch_use_1967}.

To compare charge noise data between different devices and device types, the power spectral density is fitted to:  

\begin{equation}
    S(f) \propto \frac{S_{0}}{f^{\gamma}},
\label{eq:Sf_form}
\end{equation}

\noindent where $S_0$ is the power spectral density at $f$ = 1~Hz and $\gamma$ gives an indication of the frequency dependence of the charge noise. The median value of $\gamma$ is found to be close to 1 for p-type devices, suggesting that, in general, they follow the expected trend $1/f$. 
The measured $S_0$ values are higher than typical literature values for other platforms which can range from 0.1 - 4~$\mathrm{\mu}$eV/$\sqrt{\mathrm{Hz}}$ suggesting that tackling charge noise is a critical step to allow the development of qubits in 22FDX\circledR.

\begin{table}[htbp]
\centering
\begin{tabular}{ccc}
\hline
\multicolumn{1}{|c|}{\textbf{Parameter}}                                     & \multicolumn{1}{c|}{\textbf{Median}} & \multicolumn{1}{c|}{\textbf{Interquartile Range}} \\ \hline
\multicolumn{1}{|c|}{S$_0$ ($\mathrm{\mu}$eV/$\mathrm{\sqrt{\mathrm{Hz}}}$)} & \multicolumn{1}{c|}{6.5}             & \multicolumn{1}{c|}{17.5}                         \\ \hline
\multicolumn{1}{|c|}{$\mathrm{\gamma}$}                                      & \multicolumn{1}{c|}{1.25}            & \multicolumn{1}{c|}{0.72}                         \\ \hline
\multicolumn{1}{l}{}                                                         & \multicolumn{1}{l}{}                 & \multicolumn{1}{l}{}                             
\end{tabular}
\caption{Charge noise parameters extracted from peak tracking measurements of p-type devices. \label{tbl:CN_Values}}
\end{table}

\section{Conclusions}

The single-hole transistor is the simplest device that can be used to begin the benchmarking of hole QDs in a manufacturing process. In this work, we show the classical characterization of a farm of such devices at deep cryogenic temperatures (T~<~1~K), demonstrating increased drain induced barrier lowering at shorter gate lengths and also increased normalized variability in the threshold voltage for these shorter gate devices. These findings illustrate the impact of short-channel effects and the sensitivity to manufacturing variability of devices with short-channel length when operating in the deep cryogenic regime. QD characterization is then performed, primarily using Coulomb diamonds. Using the Diamondsky package \cite{thomas_rapid_2023}, we analyzed p-type QDs as a function of the device parameters and found the device dimensions that provide a higher chance to form a well-defined QD. Overall, our results demonstrate the integration of quantum and classical electronics functioning at deep cryogenic environments and provide an efficient method to extract QD parameters at scale. Further work should concentrate on tracing the sources of variability and noise back to the manufacturing process.  

\section*{Acknowledgments}
The authors acknowledge Charan Kocherlakota, Debargha Dutta, James Kirkman, and Jonathan Warren of QuantumMotion for their technical support. We also acknowledge Nigel Cave from GlobalFoundries for useful discussions. T. S. acknowledges the Engineering and Physical Sciences Research Council (EPSRC) through the Centre for Doctoral Training in Delivering Quantum Technologies [EP/S021582/1]. A.G.-S. acknowledges an Industrial Fellowship from the Royal Commission for the Exhibition of 1851. M. F. G.-Z. acknowledges a UKRI Future Leaders Fellowship [MR/V023284/1].

\bibliographystyle{IEEEtran}
\bibliography{example}

\end{document}